\documentclass[preprint,aps,showpacs,nofootinbib,preprintnumbers,amsmath,amssymb]{revtex4-1}
\usepackage{amssymb}
\usepackage{epsfig}
\usepackage{graphicx}
\usepackage{subfigure}
\usepackage{dcolumn}
\usepackage{bm}
\usepackage[usenames ,dvipsnames]{xcolor}
\usepackage{slashed}

\begin{document}

\title{Comment on  ``A three-loop radiative neutrino mass model with dark matter" [Phys. Lett. B 741 (2015) 163]}

\author{Chao-Qiang Geng$^{1,2,3}$\footnote{geng@phys.nthu.edu.tw}, Da~Huang$^{2,3}$\footnote{dahuang@phys.nthu.edu.tw} 
and Lu-Hsing Tsai$^{2,3}$\footnote{lhtsai@phys.nthu.edu.tw}}

\affiliation{$^{1}$Chongqing University of Posts \& Telecommunications, Chongqing, 400065, China\\
$^{2}$Department of Physics, National Tsing Hua University, Hsinchu, Taiwan
\\
$^{3}$Physics Division, National Center for Theoretical Sciences, Hsinchu, Taiwan 
}

\date{\today}

\begin{abstract}
We revisit the calculation of the three-loop diagrams for the radiative neutrino mass generation and consider some relevant constraints on the model recently proposed by L.~Jin, {\it et al} (2015)~\cite{Jin:2015cla}. We find that the previous approximation is inappropriate due to the neglect of some important contributions, and the benchmark point proposed can neither give rise to enough neutrino masses nor accommodate these additional constraints, such as the validity of the perturbation theory, the electroweak precision measurements, and the neutrinoless double beta decays. 
\end{abstract}

\maketitle
\noindent The smallness of neutrino masses and the existence of dark matter are the two phenomena that require physics beyond the standard model. It is more intriguing that these two physics have a common origin. In Ref.~\cite{Jin:2015cla}, Jin, Tang and Zhang proposed an interesting model which generates neutrino masses at three-loop level, trying to explicitly provide such a connection. However, in the original estimation of neutrino masses, some important contributions, {\it i.e.} Feynman diagrams (b) and (c) in Fig.~1 of Ref.~\cite{Jin:2015cla}, were neglected. Here, we revisit the calculation of neutrino masses numerically by including these additional contributions, and find that their effect on the final results is so large that the approximation used previously cannot be justified. Then, we consider some further constraints on the model, such as the validity of the perturbation theory, the electroweak precision tests (EWPTs) and the neutrinoless double beta ($0\nu\beta\beta$) decays, which have not been discussed in Ref.~\cite{Jin:2015cla}.

By explicit numerical calculations of the three kinds of Feynman diagrams in Fig.~1 of Ref.~\cite{Jin:2015cla} with the given benchmark parameters, we obtain the following neutrino mass matrix
\begin{eqnarray} 
\label{Mu}
m_{\nu} =
\left(
\begin{matrix}
0.97 & 1.97 &   0.68 \\
1.97 & 6.20 &  5.43 \\
0.68 & 5.43  & 7.91
 \end{matrix}
\right)\times 10^{-12}~~\mathrm{GeV},
\end{eqnarray}
which is already excluded by the neutrino oscillation data, {\it e.g.}, quantitatively $(m_{\nu})_{\tau\tau}$ should not be smaller than $10^{-11}$~GeV at 3$\sigma$ C.L.
Note that our calculations include the contributions from the diagrams (b) and (c), which were previously neglected in
Ref.~\cite{Jin:2015cla}. The rational behind this approximation is that the mass of the neutral scalar $\Delta^0$, $m_{\Delta^0} = 4331~$GeV, for the choosing benchmark point is  so heavy that it greatly suppresses the diagrams (b) and (c). However, our numerical investigation clearly shows that this argument is not appropriate. In order to quantify the effects of diagram (b) and (c), we specify the contributions to the $ee$ element from three diagrams $(m_\nu)^{(a,b,c)}_{ee} = (2.25,~ -0.35,~ -0.93) \times 10^{-12}$~GeV, respectively. Obviously, the sum of the diagrams (b) and (c) gives the same order of magnitude as the diagram (a), but with an opposite sign, which leads to a large cancellation as well as the smallness of the neutrino masses for the benchmark point. Thus, it is not valid to only consider diagram (a) when estimating the neutrino masses for the present model, as was done in the original paper.   

Now we turn to some additional constraints on the present model which were previously omitted. Firstly, note that the mass difference between the two charged scalars, $H_{1,2}^-$, can be determined to be 
\begin{equation}\label{MassDiff}
m_{H_1}^2 - m_{H_2}^2 = \lambda_7 v^2/s_{2\beta}\,,
\end{equation} 
where $s_{2\beta} = \sin(2\beta)$. For the benchmark scalar masses and mixing angle, we can obtain the absolute value of the dimensionless coupling constant $|\lambda_7| = 357$, which is too large to be allowed by the perturbation theory~\cite{Nebot:2007bc}. 

Moreover, EWPTs have already placed strong constraints to the present model. In particular, the most stringent one is given by $\Delta T \in [-0.04, ~0.12]$ at $95\%$ C.L.~\cite{Baak:2012kk}, while the extra scalars give new contributions to $T$ as follows,
\begin{eqnarray}
\Delta T = \frac{1}{4\pi s_W^2 m_W^2} [c_\beta^2 F_{\Delta^0, H_1} + s_\beta^2 F_{\Delta^0, H_2} - 2 s_\beta^2 c_\beta^2 F_{H_1, H_2}]\, ,\nonumber\\
\end{eqnarray} 
where
\begin{eqnarray}
F_{i,j} = \frac{m_i^2 + m_j^2}{2} - \frac{m_i^2 m_j^2}{m_i^2 - m_j^2} \ln \frac{m_i^2}{m_j^2}\, .
\end{eqnarray}
With the above formula, the benchmark point leads to $\Delta T = -116$, which is clearly outside the experimentally allowed range. 

\begin{figure}[ht]
\centering
\includegraphics[scale=0.4]{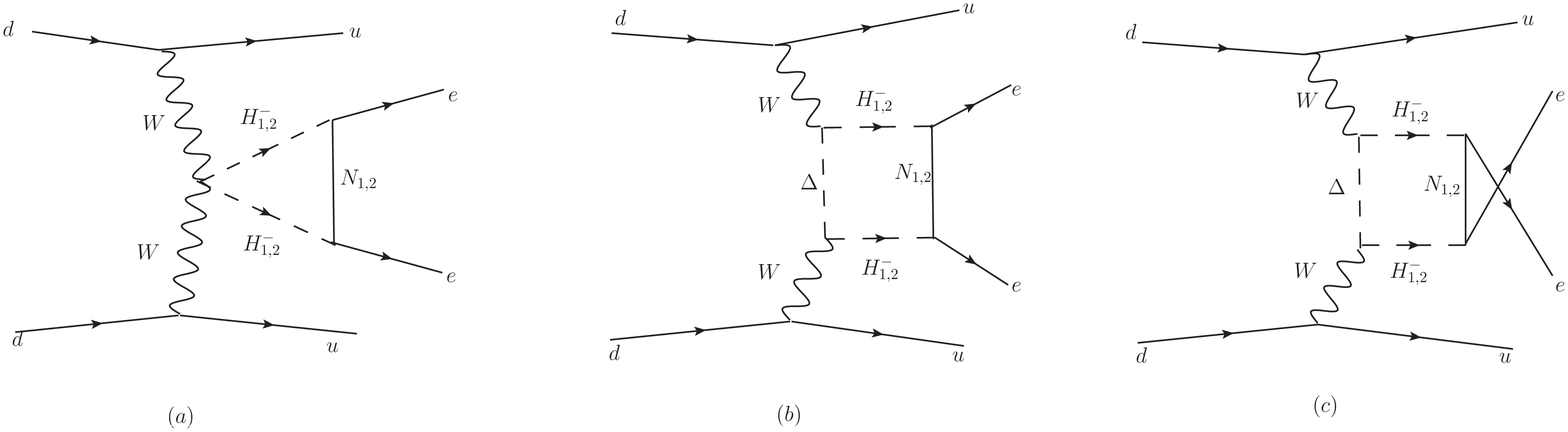}
\caption{$0\nu\beta\beta$ from short-distance channel in the present model}
\label{fig1}
\end{figure}

It is also interesting to note that this model can induce a large $0\nu\beta\beta$ decay rate via the so-called short-distance channel shown in Fig.~\ref{fig1}, which is expected to be much larger than the conventional long-distance one~\cite{Pas:2000vn,Chen:2006vn,Gustafsson:2014vpa,Geng:2014gua}.  
In the present model, such a short-distance contribution to the half lifetime for $0\nu\beta\beta$ decay is given by
\begin{eqnarray}
T^{0\nu\beta\beta} = \left[4 m^2_p G_{01} |{\cal A}|^2 |{\cal M}_3|^2\right]^{-1}\, ,
\end{eqnarray}
where $m_p$ is the mass of the proton, $G_{01}$ the phase space factor,
\begin{eqnarray}
{\cal A} = s_\beta^2 c_\beta^2 (m_{H_1}^2 - m_{H_2}^2)^2 \sum_{i=1,2} (g_{ie}^* m_{N_i} g_{ie}^* ) (I_a^i -2 I_b^i)
\end{eqnarray}
with
\begin{eqnarray}
I^i_a = -\frac{\Gamma(3)}{16\pi^2} \int^1_0 dx_1 \int^{1-x_1}_0 dx_2 \frac{x_1 x_2}{[x_1 m^2_{H_1} + x_2 m_{H_2}^2 +(1-x_1-x_2)m_{N_i}^2]^3} \, ,\nonumber\\
I^i_b = -\frac{2}{16\pi^2} \int \prod_{j=1}^3 dx_j \frac{x_1 x_2}{[x_1 m_{H_1}^2 + x_2 m_{H_2}^2 + x_3 m_{N_i}^2 + (1-x_1 -x_2 -x_3) m_{\Delta^0}^2]^3}\, ,
\end{eqnarray}
and ${\cal M}_3$ the nuclear matrix element enveloping the operator $\bar{u}_L \gamma^\mu d_L \bar{u}_L \gamma_\mu d_L \bar{e}_R e_R^c$. Given the numerical values of $G_{01}$ and ${\cal M}_3$ in Refs.~\cite{Gustafsson:2014vpa,Geng:2014gua}, we can predict the expected half lifetimes for various conventional targets. Unfortunately, with the present benchmark, the predicted half lifetimes for all the target nuclei are already excluded by the current experimental limits. The largest discrepancy comes from the measurement of $^{136}{\rm Xe}$, with the lower bound on the half life of $1.9\times 10^{25}$~yr~\cite{Gustafsson:2014vpa,Geng:2014gua}, which is compared with the calculated one of $1.1\times 10^{19}$~yr.

Finally, an interesting question is whether we can find some new benchmark point for the present model satisfying all the constraints. Note the relation $m_\nu \propto s_{2\beta}^2(m_{H_1}^2 - m_{H_2}^2)^2 = (\lambda_7 v^2)^2$. If we restrict to $\lambda_7 \leq 5$ allowed by perturbativity arguments~\cite{Nebot:2007bc} and consider the case with only two $Z_2$-odd Majorana fermions, we find that the $(m_\nu)_{ee}$ obtained is always smaller than $10^{-13}$~GeV, which is already excluded by the neutrino oscillation data with the vanishing lightest neutrino mass. From this viewpoint, we conclude that it is difficult in finding a viable benchmark point.    

\begin{acknowledgments}
This work was supported by National Center for Theoretical Sciences, National Science
Council (Grant No. NSC-101-2112-M-007-006-MY3) and National Tsing Hua University (Grant No. 104N2724E1).
\end{acknowledgments}

\end{document}